\documentclass[a4paper]{article}

\usepackage{spconf}
\usepackage{outlines}
\usepackage{subcaption}
\usepackage{xcolor}
\usepackage{makecell}

\usepackage{mparhack}   
\usepackage{lipsum}     

\usepackage{color}
\usepackage{diagbox}
\usepackage{xcolor,colortbl}
\usepackage{enumitem}
\usepackage{amssymb}
\usepackage{balance}

\makeatletter
\newcommand{\printfnsymbol}[1]{%
  \textsuperscript{\@fnsymbol{#1}}%
}
\makeatother

\usepackage[activate]{microtype}
\usepackage{times}
\usepackage{graphicx}
\usepackage{amssymb}
\usepackage{gensymb}
\usepackage{amsmath}
\usepackage{breakurl}
\usepackage{caption}
\usepackage{subcaption}
\usepackage{tikz}
\usetikzlibrary{shapes.geometric, positioning, arrows, bayesnet}

\usepackage{url,hyperref}

\usepackage{algorithm}
\usepackage{algorithmic}

\usepackage[labelfont={bf},font=small]{caption}
\usepackage[none]{hyphenat}

\usepackage{mathtools, cuted}

\usepackage[noadjust, nobreak]{cite}

\usepackage{tabularx}
\usepackage{amsmath}

\usepackage{float}

\usepackage{pifont}%

\usepackage[activate]{microtype}
\title{Detecting COVID-19 from Breathing and Coughing Sounds\\ using Deep Neural Networks}

\name{Bj\"orn W. Schuller$^{1,2}$,
Harry Coppock$^{2}$, and Alexander Gaskell$^2$}
\address{
  $^1$Chair of Embedded Intelligence for Health Care and Wellbeing, University of Augsburg, Germany\\
  $^2$GLAM -- Group on Language, Audio, \& Music, Imperial College London, UK
\texttt{\small 
}\\
  }

\begin{document}

\maketitle
\begin{abstract}
The COVID-19 pandemic has affected the world unevenly; while industrial economies have been able to produce the tests necessary to track the spread of the virus and mostly avoided complete lockdowns, developing countries have faced issues with testing capacity.
In this paper, we explore the usage of deep learning models as a ubiquitous, low-cost, pre-testing method for detecting COVID-19 from audio recordings of breathing or coughing taken with mobile devices or via the web.
We adapt an ensemble of Convolutional Neural Networks that utilise raw breathing and coughing audio and spectrograms to classify if a speaker is infected with COVID-19 or not.
The different models are obtained via automatic hyperparameter tuning using Bayesian Optimisation combined with HyperBand.
The proposed method outperforms a traditional baseline approach by a large margin. Ultimately, it achieves an Unweighted Average Recall (UAR) of 74.9\,\%, or an Area Under ROC Curve (AUC) of 80.7\,\% by ensembling neural networks, considering the best test set result across breathing and coughing in a strictly subject independent manner. In isolation, breathing sounds thereby appear slightly better suited than coughing ones (76.1\,\% vs 73.7\,\% UAR).
\end{abstract}


\section{Introduction}
\label{sec:intro}

The COVID-19 pandemic has forced the global community to recon with a shortage of adequate testing capacity all over the globe \cite{testingsurvey}. The Foundation for Innovative New Diagnostics (FIND) tracker \cite{testtracker} shows that the pandemic has exacerbated global economic inequalities. Developed countries have mature industries available with which testing equipment and materials can be produced locally or the funds to procure the necessary materials from abroad. Meanwhile, developing countries face a distinct lack of testing equipment and materials, having even during the second wave of the pandemic the capacity to only conduct less than 5\,000 daily tests per capita, or even 1\,000 tests per capita for some developing countries. This lack of access to testing capacity forces countries to grapple with two options: either push through hard or complete lockdown measures in bids to slow down or break the spread of the virus, placing excessive strain on local economies for those workers whose jobs can not be done in a Work from Home (WfH) fashion \cite{gottlieb2020lockdown}, or allow the virus to pass undetected through their populations in trying to shore up the economies, but while placing further pressure on already under-equipped health systems.

Artificial intelligence (AI) and in particular, Deep Neural Networks (DNNs), started to grow in popularity ever since amongst others \cite{NIPS2012_c399862d} used them to surpass -- by a large margin -- previous classical machine learning approaches on the ImageNet Large Scale Visual Recognition Challenge (ILSVRC) \cite{ILSVRC15}. Since then, they have come to set the state of the art in a number of different fields and challenges. Among these fields where AI has been finding usage is that of medicine in general \cite{HAMET2017S36},
and for COVID-19 in particular \cite{schuller2020covid}.
Of particular interest here, is the usage of AI to help in combating the COVID-19 pandemic, as researchers have already made use of AI to analyse different signals for signs of COVID-19 \cite{deshpande2020overview,9069255}.

It is in this context that it becomes important to consider the usage of everyday tools such as internet-connected mobile phones \cite{pena2016world}, together with artificial intelligence as tools to detect the infection with COVID-19 as a way of mitigating the necessity for everybody to take tests, or to quarantine at home.
In this paper, we introduce a machine learning model that detects, by way of coughing or breathing samples, whether the audio contains traces of COVID-19 or not,
which could be used, e.\,g., as pre-selection filter for more reliable, but more expensive testing methods. The audio stems from a public database of coughing and breathing sounds collected from mobile phones and over the Internet crowd-sourced by the University of Cambridge.

For an overview on the current state-of-play in COVID-19 detection from audio, a short overview is given in \cite{Qian20-RAI}.

In Section \ref{sec:approach} we introduce our approach, including how we represent the audio as input to the models, the model architecture and construction, the training technique we used, as well as the hyperparameter optimisation process undertaken; we finish the section by introducing the baseline with which we compare our results. In Section \ref{sec:experiments}, we showcase our experiments, beginning by introducing the dataset that serves as the basis for our experiments, the evaluation metrics that we consider, and the results of our approach, as well as finish the section with a discussion of the results. We conclude the paper in Section \ref{sec:conclusion}.

\section{Approach}
\label{sec:approach}

Our approach utilises raw audio in combination with different spectrogram variations. 
We implement an independent branch in our model for each of the input formats, where each branch consists of a Convolutional Neural Network (CNN).
Eventually, we combine the learnt representations from each branch by concatenation and use fully-connected layers to give a final classification label.

\subsection{Models}
\label{subsec:models}

Our models make use of two building blocks. We will introduce each of the building blocks first, followed by an explanation of the general architecture of the models.
We refer to a \emph{temporal feature map} as a real-valued 2-dimensional tensor of dimensions $T \times F$, where $T$ is the length of time dimension and $F$ is the number of features. According to this definition, the spectrograms and raw audio are both temporal feature maps.

The first building block is a \emph{convolution block}, which consists of a 1D convolution layer with $C$ output channels and Rectified Linear Unite (ReLU) activation function, followed by a max pooling layer, then a dropout layer \cite{dropout}. The dropout layer has the aim of reducing overfitting; we adapt a dropout rate of $0.2$.
Additionally, a \emph{convolution branch} is a stack of $N$ convolution blocks.
A convolution branch will map a temporal feature map into another temporal feature map, which is more condensed in the time dimension. In other words, the features at the end have high representations, while the time dimension is reduced several times via the the pooling layers in the convolution blocks.

The second building block is a reduction of the time dimension of a temporal feature map, using global pooling layers.
We make use of global average pooling and global maximum pooling; then, we concatenate the resulting features from both.

For the model, we make use of several representations of the input, inspired by \cite{Szep2020} which utilised multi-channel spectrograms.
The spectrograms are grouped by their hop-length, because different hop-lengths result in different lengths of the time dimension of the spectrograms.
Each group of spectrograms is concatenated along the channels axis, which results in a total of two groups of spectrograms in addition to the raw audio.
For each of those three, we construct separate but identical convolution branches, followed by the global pooling. After that, the outputs from the different branches are concatenated, and followed by $F$ fully-connected layers, which include the last layer the outputs the classification labels.
%

\subsection{Training and hyperparameter optimisation}
\label{subsec:training}

We train several variants of the model architecture described in Subsection~\ref{subsec:models} while splitting the data using nested $k$-fold cross validation \cite{nestedkfold} on the mixture of train and development parts of the data. 
The variants differ in at least one of two aspects; they are either trained using a different fold or constructed having different hyperparameters.
We utilise an ensemble of the variants to reach best results, as ensembling tends to reduce overfitting and hence reaches the best results \cite{friedman2001elements}. We use averaging of the prediction probabilities to ensemble the different models. 
For all the variants, we employ the Adam optimiser \cite{Adam} using a mini-batch size of $16$.
The learning rate of Adam $\alpha$ is considered as a hyperparameter to be tuned.
Since the database consists of audio tracks of varying lengths, we pad the examples with 0 values to match the longest example in a given batch.
The network parameter optimisation is done by minimising the binary crossentropy loss function, while giving a different weight $\lambda$ to the positive class. This will direct the model to focus more on getting the positive cases correct, which is helpful in case the positive cases are underrepresented in the data. The final equation of the loss function is given by:
\begin{equation*}
\label{eqn:cross}
    \mathcal{L}(\textbf{y}, \hat{\textbf{y}}) = -\frac{1}{N} \sum_{i=1}^N {(\lambda y_i \log \hat{y}_i + (1 - y_i) \log (1 - \hat{y}_i))}.
\end{equation*}
For hyperparameter tuning, we use Bayesian Optimisation HyperBand (BOHB) \cite{BOHB} by instructing it to reach the best Unweighted Average Recall (UAR) (explained further in Subsection \ref{subsec:metrics}) on the validation data, namely by maximising the average of the UAR across the hold out folds in the cross-validation. 

\subsection{Baseline}
\label{subsec:baseline}
We compare our approach against a baseline linear Support Vector Machine (SVM) classifier \cite{bishop2006pattern}, which is trained on audio functionals extracted with openSMILE~\cite{eyben2010opensmile}.
We choose the large feature configuration (6,373 functionals) as introduced in the Interspeech 2016 ComParE challenge~\cite{schuller2016interspeech}, which has been the configuration of choice for the Interspeech ComParE challenges ever since.
On that, we perform Principal Component Analysis (PCA) \cite{bishop2006pattern} to reduce the number of features to 100 in a standard manner.
The SVM classifier's complexity parameter is optimised on a logarithmic scale between $[10^{-5}, 1]$, based on the achieved UAR on a development partition. The optimisation yielded a value $10^{-5}$.
Afterwards, the SVM is fit on the combined training and development data with the optimised complexity value and finally used for inference on the test partition of the data.

\section{Experiments}
\label{sec:experiments}
\begin{table}[]
    \centering
    \begin{tabular}{l|l|l|r}
         Condition &  Platform & Symptom & \# files\\
         \hline
         Asthma & Android & cough & 26 \\
         Asthma & Web & cough & 16 \\
         
         COVID & Android & no cough & 128\\
         COVID & Android & cough& 92\\
         COVID & Web & no cough& 46\\
         COVID & Web & cough & 16\\
         
         Healthy & Android & no symptom& 282\\
         Healthy & Android & cough& 16\\
         Healthy & Web & no symptom& 362\\
         Healthy & Web & cough & 50\\
         
    \end{tabular}
    \caption{Unaugmented database file statistics}
    \label{tab:data_stats}
\end{table}
\begin{table*}[!t]
    \centering
    \begin{tabular}{c|c|c|c|c|c|c||c||c|c|c|c|c}
        \multicolumn{7}{c||}{Model description} & Valid. & UAR & AUC & ACC  & ACC$_{+}$ & ACC$_{-}$ \\ \hline
            \hline 
            \multicolumn{7}{c||}{Baseline (openSMILE ComParE features + SVM)}  & - & 59.3 & 61.0 & 56.6 & 67.2 & 51.3 \\
            \hline
            \hline
        index & $N$ & $C$ & $F$ &  $\lambda$ & $\alpha$ & Fold &&&&&&\\ 
            \hline

            1  &  5  &  77  &  2  &  1.582  &  $4.8 \times 10^{-5}$  &  1  &  65.8  &  71.4  &  76.0  &  72.6  &  68.1  &  74.8 \\
            2  &  2  &  308  &  2  &  2.093  &  $5.9 \times 10^{-6}$  &  1  &  61.4  &  66.0  &  74.8  &  59.4  &  85.3  &  46.6 \\
            3  &  2  &  235  &  2  &  1.805  &  $2.6 \times 10^{-5}$  &  1  &  60.7  &  64.6  &  72.8  &  62.3  &  71.6  &  57.7 \\
            4  &  3  &  466  &  2  &  1.859  &  $7.1 \times 10^{-6}$  &  1  &  60.6  &  63.4  &  71.1  &  59.7  &  74.1  &  52.6 \\
            \hline
            5  &  4  &  73  &  2  &  1.917  &  $9.7 \times 10^{-6}$  &  2  &  63.8  &  53.4  &  59.6  &  52.3  &  56.9  &  50.0 \\
            6  &  5  &  151  &  1  &  2.176  &  $1.8 \times 10^{-6}$  &  2  &  62.1  &  57.7  &  73.1  &  44.9  &  95.7  &  19.7 \\
            7  &  3  &  486  &  1  &  1.678  &  $8.8 \times 10^{-6}$  &  2  &  61.4  &  60.5  &  65.3  &  58.6  &  66.4  &  54.7 \\
            8  &  5  &  435  &  1  &  1.351  &  $3.7 \times 10^{-5}$  &  2  &  59.9  &  69.1  &  72.9  &  68.3  &  71.6  &  66.7 \\
            \hline
            9  &  4  &  445  &  2  &  0.748  &  $2.3 \times 10^{-5}$  &  3  &  63.2  &  58.5  &  64.8  &  50.9  &  81.0  &  35.9 \\
            10  &  2  &  215  &  2  &  1.218  &  $7.3 \times 10^{-6}$  &  3  &  63.2  &  52.4  &  60.4  &  52.3  &  52.6  &  52.1 \\
            11  &  3  &  486  &  1  &  0.882  &  $4.3 \times 10^{-6}$  &  3  &  61.3  &  52.9  &  59.3  &  61.4  &  27.6  &  78.2 \\
            12  &  6  &  424  &  1  &  1.328  &  $2.3 \times 10^{-5}$  &  3  &  60.6  &  70.4  &  74.8  &  68.9  &  75.0  &  65.8 \\

            \hline
            \hline
            \multicolumn{5}{l|}{Ensemble description} & \multicolumn{2}{c||}{Ensembled group}  & &&&&& \\
            \hline
            
            \multicolumn{5}{l|}{Best test UAR} & \multicolumn{2}{c||}{1,2,6,7,8,11}  &  -  &  \textbf{74.9}  &  80.5  &  \textbf{73.1}  &  80.2  &  69.7  \\
            \multicolumn{5}{l|}{Best test AUC} & \multicolumn{2}{c||}{1,2,6,7,8,11,12}  &  -  &  74.5  &  \textbf{80.7}  &  72.9  &  79.3  &  69.7  \\
            \multicolumn{5}{l|}{Best model per fold} & \multicolumn{2}{c||}{1,5,9}  &  -  &  70.8  &  77.3  &  68.3  &  78.4  &  63.2    \\
            \multicolumn{5}{l|}{All} & \multicolumn{2}{c||}{1-12}  &  -  &  70.2  &  77.6  &  67.7  &  77.6  &  62.8  \\

    \end{tabular}
    \caption{Results of the different models using the introduced performance measures in percentages: validation UAR on hold out set, followed by test set UAR, AUC, ACC, and class-wise (+/-) ACC (recall/sensitivity), respectively.
    The models are the baseline, several CNNs with different hyperparameters or training folds, and ensembled models. $N, C, F$ are model hyperparameters (Subsection~\ref{subsec:models}), and $\lambda, \alpha$ are training hyperparameters (Subsection~\ref{subsec:training}). The models used for the ensemble groups are specified by the index column. The best model per fold is specified by the best validation UAR on the hold out set.
    }
    \label{tab:results}
\end{table*}

\begin{table}[]
    \centering
    \begin{tabular}{c|c|c}
    Model summary mean $\pm$ std. & UAR & AUC \\
    \hline
    \hline
    Fold 1 models & 66.4 $\pm$ 3.1 & 73.7 $\pm$ 1.9 \\
    Fold 2 models & 60.2 $\pm$ 5.7 & 67.7 $\pm$ 5.6 \\
    Fold 3 models & 58.6 $\pm$ 7.2 & 64.8 $\pm$ 6.1 \\ \hline
    Ensembled models & 72.6 $\pm$ 2.1 & 79.0 $\pm$ 1.6
    \end{tabular}
    \caption{An overview on result uncertainties by summary statistics of the UAR and AUC mean and standard deviations in percentages for the models from each individual fold , as well as for all the ensembled models reported in Table \ref{tab:results}}
    \label{tab:meanstd}
\end{table}

\begin{table}[]
    \centering
    \begin{tabular}{c|c|c}
      Ensemble   & breath UAR & cough UAR\\
      \hline \hline
      1,2,6,7,8,11  &  \textbf{76.1}  &  \textbf{73.7}\\
       1,2,6,7,8,11,12  &  75.7  &  73.3 \\
       1,5,9  &  71.0  &  70.7 \\
      1-12   &  71.4  &  69.0 \\ \hline
      Ensembles mean $\pm$ std. & 73.6 $\pm$ 2.4 & 71.7 $\pm$ 1.9\\
    \end{tabular}
    \caption{The breath and cough UAR percentages on the test samples with the corresponding modality only.}
    \label{tab:cough}
\end{table}  

\subsection{Database}
\label{subsec:database}

The experiments were performed on a crowdsourced database \cite{brown2020exploring} that was collected via the ``COVID-19 sounds'' Android app, as well as through a web form (an iOS app also exists, though it did not contribute to the database used here) by the University of Cambridge \cite{brown2020exploring}. The participants are asked to fill a survey about their demographic information (such as age and location), medical history, as well as symptoms (if any). The app instructed the participants to ``breathe deeply five times, cough three times, and read three times a short sentence appearing on screen'' \cite{brown2020exploring}.
\footnote{Please refer to https://www.covid-19-sounds.org/en/app/ for more details.}


The present database is an excerpt of the total data collected via the app. The database includes the coughing and breathing audio samples, their associated medical condition (COVID, Asthma, or Healthy), whether the submitter suffered from coughing as a symptom or not, and finally which platform the samples were collected through (Android or Web).

The database consists of a total of 1\,427 audio files that total 3.93 hours. Of these 1\,427 files, 1\,034 are original audio samples that total two hours, while the rest are the result of data augmentation \cite{jan_schluter_2015_1417745}. The mean file length for the dataset is 9.96 seconds with a standard deviation of 6.02 seconds. The audio files have a sample rate 16\,kHz and 16 bit quantisation. In total, 174 unique participants are included in the database. 

For the purposes of our experiments, only the non-augmented files are used. The audio files are downsampled to 8\,kHz and the different database labels resembling telephone speech quality in this respect and are collapsed into two sets of experiment labels: \emph{covid positive}, which includes all the COVID-19 labels, while all the other label categories are classified as \emph{covid negative}. The summary statistics of the labels of the unaugmented audio files can be found in Table \ref{tab:data_stats}.

We split the data in a stratified, strictly subject-independent manner into three roughly balanced sets: train, dev(elopment), and test. The train set consists of 464 samples, the dev set consists of 220 samples, and the test set consists of 350 examples. The models in our experiments are trained using $k$-fold nested cross validation \cite{nestedkfold},  
with $k=3$, wherein each model is trained on two thirds of the mixture of the train and dev sets, with its performance during training evaluated on the remaining hold out part\footnote{The split indices are available by the authors for reproducibility.}.
We saved models with best Unweighted Average Recall (UAR) on the hold out set.

In other words, we mix the train and development, and then split this speaker independently into three folds. Then, we train different models on different folds, and we test them at the end on the overall `final' test set, which is always strictly held out also in case of model fusions. During training, we validate on the hold out fold and never look at the test set.
Likewise, all four parts: Fold1, Fold2, Fold3, and test are pairwise mutually exclusive in terms of speakers, so a speaker appears only in one of these 4. Fold1 is what we declare as the dev(elopment) part for the baseline. Note that, as the splits are generated in a stratified manner, in all splits, the COVID-19/total ratio is close to the ~27\,\% of the COVID-19/total ratio of the whole dataset.

Also, one can see that on some folds, high validation UAR yields bad test UAR and vice versa, which shows we are not biasing our training for getting good test results.

\subsection{Evaluation metrics}
\label{subsec:metrics}
We adapt several classification metrics:
\begin{itemize}[leftmargin=*]
    \item Unweighted Average Recall (UAR): Also known as Unweighted Average Accuracy, is the sum of class-wise accuracy (recall) divided by number of classes -- the standard competition measure in the Interspeech ComParE challenge series. Chance level for two classes resembles 50.0\,\% UAR.
    \item Area Under ROC Curve (AUC): The ROC curve plots the true-positive rates against the false-positive rates with varying classification thresholds, and measures the area under the drawn curve. This measures how well can a model distinguish between the true and false classes, where a random baseline classifier will get a value of 50.0\,\%.
    \item Accuracy (ACC): The ratio of examples that are answered correctly in the evaluation set.
    \item Recall COVID-19 positive (ACC$_{+}$).
    \item Recall COVID-19 negative (ACC$_{-}$).
\end{itemize}

UAR is a preferred measure as compared to the Accuracy, because the Accuracy will reward trivial classifiers on unbalanced data, where the classifier always just predicts the more frequent class (here, the negative class).

\subsection{Results and discussion}
\label{subsec:results}

In Table~\ref{tab:results}, we showcase the following models: The baseline model, trained using openSMILE ComParE features (introduced in Subsection \ref{subsec:baseline}), individually trained models (introduced in Subsection \ref{subsec:models}), whose hyperparameters are tuned using BOHB.  
The range for the parameters examined is  $N$ in [2,6], $C$ in [64,512], $F$ in [1,2] (number of fully connected layers at the end, including the final classification layer), $\alpha$ in [$10^{-7}$,$10^{-3}$], and log sampled $\lambda$ in [0.5,2.2]. 
Note that, if lambda is less than one, the model will focus on the negative class, which non-intuitively yielded good results in few cases.
Finally, we present the ensemble models (introduced in Subsection \ref{subsec:training}). The performance of the models is shown using the metrics introduced in Subsection \ref{subsec:metrics}. The metrics themselves indicate performance on the test set, while the validation column indicates the performance of the individual models on the hold-out set of their particular fold.

As can be seen in Table \ref{tab:results}, all the variations of ensembled models outperform the baseline by a wide margin on all metrics. The same cannot be said for the individual models trained on one of the three total folds: while all of the models trained on the first fold outperform the baseline for all the metrics, only two to three of the models trained on folds two and three surpass the baseline, depending on the metric. That being said: model six produces the best overall positive class recall score of 95.7$\,\%$, while model eleven does the same for the negative class recall, with a score of 78.2$\,\%$, with both of them surpassing the respective metric scores of any of the ensembles, yet, athe cost of the respective other class. Still, these results probably justify why both models were included in both of the best performing ensembles.
%

In terms of ensembles, we show the following configurations: An ensemble that provides the best test set UAR, an ensemble that provides the best test set AUC, an ensemble of the best model per fold (based on the UAR metric) and finally, an ensemble of all the trained models. As would be expected, assembling a particular set of models based on which combination reaches the best test set performance for a particular metric does indeed yield the highest result for that metric. For UAR, the highest result is 74.9$\,\%$, while for AUC, the highest result is 80.7$\,\%$. Selecting only the best model per fold and ensembling those together results in  performances of 70.8$\,\%$ UAR and 77.3$\,\%$ AUC. Finally, ensembling all models together reaches the results of 70.2$\,\%$ UAR, and 77.6$\,\%$ AUC.

In order to provide an insight into uncertainty of the results, in Table \ref{tab:meanstd}, we pick the UAR and AUC metrics for deeper inspection and calculate the mean and standard deviations for the respective metrics of the models produced from each of the folds, as well as those for the ensembled models. The table shows that the models trained on fold 1 are on average much more capable than their equivalents trained on folds 2 and 3. The table shows that not only are the means of the UAR and AUC metrics for fold 1 models around 6$\,\%$ higher than the next best averages, but they are also much more consistent, having smaller standard deviations as well. The models trained on folds 2 and 3 suffer from higher standard deviations, with the strength of the models trained on these folds being apparently dependent on random factors related to model structure or even parameter initialisation. In contrast, standard deviation for the ensembled methods is lowest both for UAR and AUC.

In Table \ref{tab:cough}, we show the UAR performances of the ensembles, as well as the summary statistic mean and standard deviation on the two different collected audio modalities (breath and cough) separately. The table shows that even with the different modalities reported separately, the best overall test set UAR ensemble gives the best UAR results for each of the modalities. The table further shows that breathing seems to be better suited than coughing.

From Table \ref{tab:results}, it can seen that it appears indeed possible to use cough and breathing audio samples, fed to neural network models to predict whether a patient has COVID-19 or not, e.\,g., for a pre-diagnosis to pre-select candidates for more reliable, yet more effort and cost requiring testing. The results shown above should be considered as initial results, given the limited size of the data set of two hours of original samples, which prevents us from using larger, more complex neural networks. The effects of the small size of the dataset can be seen in Table \ref{tab:meanstd}, which shows that the specific selection of which fold a model is trained on impacts the predictive strength of the model, a factor that can be reduced by increasing the data size, and thus allowing for fold splits with more uniform information content. 
Table \ref{tab:cough} shows that the UAR performances for the breath modality are consistently higher, suggesting that breath audio samples contain more COVID-19 information, or at the very least information that is easier to extract. These results would suggest that future data collection efforts should place a particular focus on collecting more breathing audio recordings, in addition to more validated COVID-19 test results, as well as a more diverse range of control group samples from other respiratory diseases. 




\section{Conclusion}
\label{sec:conclusion}

In this paper, we explored the usage of deep learning models as a way to predict whether someone is infected with COVID-19 based on an audio sample of either their breathing or their coughing.
The need for this usage arose from issues with lacking COVID-19 testing capacity in developing countries across the world, as opposed to the abundance of mobile phone quality microphones, but also the general opportunities coming with real-time low cost pre-scanning for selective testing with more reliable approaches.
Accordingly, the aim of the models would be to function as a ubiquitous, low-cost pre-testing mechanisms that could help mitigate the demand for COVID-19 lab tests, which are relatively expensive to conduct, as they require access to materials, equipment and manpower that are not equally available around the world.

To this end, we used a subset of a crowdsourced database collected via the University of Cambridge's COVID-19 Speech Android app and web interface. The database contained samples of breathing and coughing recordings, as well as associated demographic information, medical history, and COVID-19 testing status.
We illustrated how we pre-processed the database, splitting it in a stratified, strictly subject-independent manner into 3-fold train and development sets, as well as an independent test set. We then showed how we trained a number of individual Convolutional Neural Networks (CNNs), which we then ensembled together in order to produce our predictions.
Our proposed models achieved at best a UAR score of $74.9\,\%$ and an AUC score of $80.7\,\%$ on the held-out speaker independent test partition. 

The achieved results suggest that it is indeed possible to detect COVID-19 by way of either breath or cough samples with an accuracy relevant to use-cases such as pre-selection for more reliable testing, and also possible to use deep learning models to perform this detection. However, the current results are limited by amount of available data, which might prevent the usage of even larger models, which is where deep learning models tend to produce their best results. A future direction for this research would be to collect a larger database with highly validated and more varied control data, including a plethora of other respiratory and further related diseases, which would open the door to even better, but also more tangible results.



\section{Acknowledgement}
The authors dedicate their utmost special thanks to their colleagues Mina A.\ Nessiem and Mostafa M.\ Mohamed for their great help.
The University of Cambridge does not bear any responsibility for the analysis or interpretation of the data used herein, which represents the own view of the authors. 

\bibliographystyle{IEEEtran}
\bibliography{refs}

\end{document}